\documentclass[12pt]{report}
\usepackage{graphicx}

\def\mbi#1{\mbox{\boldmath$#1$}}
\def\kp{\mbi{k} \cdot \mbi{p}}

\def\beeq{\begin{equation}}
\def\eneq{\end{equation}}
\def\beeqa{\begin{eqnarray}}
\def\eneqa{\end{eqnarray}}

\addtocounter{section}{-1}
\begin{document}

\begin{center}

\vspace{2cm}

{\large {\bf {Theoretical study on novel electronic properties\\
in nanographite materials
} } }

\vspace{1cm}

{\rm Kikuo Harigaya$^{1,2,}$\footnote[1]{Corresponding author;
FAX: +81-29-861-5375; E-mail: k.harigaya@aist.go.jp},
Atsushi Yamashiro$^1$, Yukihiro Shimoi$^{1,2}$,
Katsunori Wakabayashi$^3$,
Yousuke Kobayashi$^4$, Naoki Kawatsu$^4$, 
Kazuyuki Takai$^4$, Hirohiko Sato$^5$,\\
J\'{e}r\^{o}me Ravier$^4$, Toshiaki Enoki$^4$, and
Morinobu Endo$^6$}

\vspace{1cm}

{\sl $^1$Nanotechnology Research Institute, AIST, 
Tsukuba 305-8568, Japan\\
$^2$Synthetic Nano-Function Materials Project, 
AIST, Tsukuba 305-8568, Japan\\
$^3$Dept. of Quantum Matter Science, 
Hiroshima University, Higashi-Hiroshima 739-8530, Japan\\
$^4$Department of Chemistry, 
Tokyo Institute of Technology, Meguro-ku 152-8551, Japan\\
$^5$Department of Physics, Chuo University, 
Bunkyo-ku 112-8551, Japan\\
$^6$Faculty of Engineering, Shinshu University, 
Nagano-shi 380-8553, Japan}

\end{center}

\vspace{1cm}

\noindent
{\bf Abstract}\\
Antiferromagnetism in stacked nanographite is investigated with 
using the Hubbard-type model.  We find that the open shell 
electronic structure can be an origin of the decreasing magnetic 
moment with the decrease of the inter-layer distance, as 
experiments on adsorption of molecules suggest. Next, possible 
charge-separated states are considered using the extended 
Hubbard model with nearest-neighbor repulsive interactions.  The 
charge-polarized state could appear, when a static electric field 
is present in the graphene plane for example.  Finally, 
superperiodic patterns with a long distance in a nanographene 
sheet observed by STM are discussed in terms of the interference 
of electronic wave functions with a static linear potential 
theoretically. In the analysis by the $\kp$ model, the oscillation 
period decreases spatially in agreement with experiments.

\vspace{1cm}

\section{Introduction}

Nanographite systems, where graphene sheets of the orders of the 
nanometer size are stacked, show novel magnetic properties, such 
as, spin-glass like behaviors \cite{1}, and the change of ESR line 
widths while gas adsorptions \cite{2}.  It has been found \cite{3,4} 
that magnetic moments decrease with the decrease of the interlayer 
distance while water molecules are attached physically. Such the 
phenomena indicate the tunable magnetism in nanometer size systems. 
Recently, unique magnetisms in carbon-based materials have been 
observed \cite{5,6}, and thus theoretical investigations are urged 
in order to resolve the mechanisms.

This paper reviews recent theoretical works on electronic
properties in nanographite materials, and is organized as follows.
(1)  Antiferromagnetism in the stacked nanographite is 
investigated with using the Hubbard-type model taking into 
account of the hopping interactions of $\pi$-electrons and the 
strong onsite repulsions at carbon atoms \cite{7,8}. We point out
the importance of the open shell electronic structure as an origin 
of the decreasing magnetic moment with the decrease of the 
inter-layer distance \cite{3,4}.  
(2) Possible charge-separated states in nanographite
ribbons are discussed in terms of the extended Hubbard model
with nearest-neighbor interactions \cite{9}.  Such the 
charge-polarized state could be observed, when a static electric
field is applied in the direction parallel with the graphene 
plane for example.  
(3) Superperiodic patterns with a long distance in a nanographene 
sheet observed by STM are discussed in terms of the interference 
of electronic wave functions. The period and the amplitude of 
the oscillations decrease spatially in one direction. We 
explain the superperiodic patterns with a static linear 
potential in the $\kp$ model \cite{10}.

\section{Tuning magnetism in stacked nanographite}

First, we report magnetic properties for the A-B stacked 
hexagonal nanographite shown in Fig. 1 (a) \cite{7}.
The first and second layers are displayed by the thick and
thin lines, respectively.  In each layer, the nearest neighbor
hopping $t$ is considered. Each layer has closed shell electron
systems when the layers do not interact mutually, because the 
number of electrons is even and equal to the number of sites.  The 
interlayer hopping $t_1$ is assigned at the sites 
with closed circles.  The model is solved with the unrestricted
Hartree-Fock approximation, and antiferromagnetic solutions
are obtained.  Figure 1 (b) shows the absolute value of 
the total magnetic moment per layer as functions of $t_1$ and $U$.
As increasing $U$, the magnitude of the magnetization increases.  
The magnetic moment is zero at the smaller $t_1$ region 
for $U=1.9t$ (open squares), $2.0t$ (closed circles), and $2.1t$
(open circles).  The magnetic moment is zero only at $t_1 = 0$
for $U=2.2t$ (closed triangles) and $2.3t$ (open triangles).  
We can understand the parabolic curves as a change due to the 
Heisenberg coupling proportional to $t_1^2/U$.

Next, we consider systems with open shell electronic structures 
when a nanographene layer is isolated \cite{8}. One case is the effects 
of additional charges coming from functional side groups with 
introducing a site potential $E_s$ at edge sites.  We take 
$E_s = -2t$, and one additional electron per layer is taken 
account. Figure 2 displays the absolute values of total magnetic
moment per layer.  In Fig. 2 (a), the site potentials 
locate at the site A in the first layer [Fig. 1 (a)], 
and at its symmetrically equivalent site in the second layer.  
In Fig. 2 (b), the site potential exists at the site B.
The total magnetization is a decreasing function of $t_1$
in both figures.  The decrease is faster in Fig. 2 (b)
than in Fig. 2 (a).  The site B is neighboring to the
site with the interaction $t_1$, and thus the localized
character of the magnetic moment can be affected easily
in this case.  The decease of observed magnetization by the
magnitude $30-40$\% with the water molecule attachment \cite{3,4}
may correspond to the case of Fig. 2 (b). 
The other origin, i.e., the geometrical effects can give rise 
to the decrease of the magnetic moment with increasing interlayer 
hopping interactions, too.  Therefore, we conclude that the open 
shell nature of electronic states in a single graphene layer 
plays an important role in controlling the magnetism in 
nanographite materials.

\section{Charge- and spin-separated states in nanographite ribbons}

In section 2, we have discussed the magnetic properties with using
the onsite interactions $U$ only.  In this section, we will report
a possible charge-separated state in nanographite ribbons
with zigzag edges by introducing the nearest-neighbor Coulomb repulsion 
term $V$ as well as $U$ \cite{9}.  The model is solved with
a mean field method with the finite space geometry shown in Fig. 3.
Charge- and spin-polarized solutions are obtained depending on the
interaction strengths.  For example, Fig. 3 (a) shows the charge
density distribution of the ferroelectric charge separated (CS) 
state with $U = 0.3t$ and $V=0.4t$. This state appears and becomes 
stable when the effects of $V$ overcom those of $U$.  
On the other hand, Fig. 3 (b) shows the spin density profile 
of the localized magnetic (LM) state for $U=t$ and $V=0$. 
This state is stable when $V$ is weak.  The CS (LM) state has no 
spin (charge) density at every site. In the CS state, the upper 
(lower) zigzag edge is charged positively (negatively). This 
distribution pattern is quite similar to that in the LM state.  
Such the static charge polarization (magnetization) is explained 
by the interplay of the Fermi instability of the flat bands due 
to $V$ ($U$) and the localized edge states.  The signs of the 
charge (spin) densities at neighboring sites are opposite, 
reflecting the bipartite nature.

Figure 4 shows the phase diagram in the parameters $U$ and $V$,
representing the stability between the CS and LM states.  We use 
the geometry with $4{\times}40$ carbon atoms.  Above the phase 
boundary, the CS state has the lower energy, and the LM state 
becomes stable below the boundary.  The phase boundary indicates
the first order phase transition.  At $U=0$, the phase boundary 
rises up with the infinite gradient. This is a signature of 
the localized edge states, differently from the graphite sheet.
The inset shows the phase diagram in reduced scales. 
In the strong correlation limit, the phase boundary approaches 
to  an  analytical  phase boundary line $V = [N/(3N-1)] U$ 
between the CDW and SDW states, which is obtained by equalizing 
their Coulomb energies.  Here, $N$ is the number of zigzag lines
of the ribbon.  The CDW and SDW solutions are crossovered 
from those of CS and LM for the weak interactions.
Such the charge-separated states could be observed
when static electric field is applied in the nanographene
plane for example, and the presence of the state will give 
effects on dielectric properties.

\section{Electronic wave interference patterns}

In Fig. 5, an STM image of the graphene sheet with a necktie 
shape is shown \cite{10}. The observation has been done with the following 
condition: bias voltage $V = 200$ mV and current $I = 0.7$ nA. The 
distance between the graphene necktie and the substrate is over 
0.8 nm, suggesting that it consists of a stacking of two graphene 
layers, which interact weekly with the HOPG substrate. 
Interestingly, the period and the amplitude of the oscillations 
decrease from the top to the bottom along the graphene necktie. 
The oscillation period is one order of magnitude larger than that 
of the Moir\'{e} pattern due to stacking, which has been reported 
elsewhere, and therefore this possibility may be weak. We can 
assume effects of long-distance periodic-structural deformations 
in the graphene surface or interference effects of electronic 
wave functions. We have also observed that the oscillation 
period becomes longer by placing a nanographene flake on the 
graphene necktie. The oscillations period seems to be double 
in the upper region of the necktie after addition of one flake. 
The oscillation below the flake seems to be only slightly 
modified by the flake. Such effect on the oscillations cannot 
be explained by some structural modulations. Therefore, the 
oscillation patterns could be the effect of interference of 
the electronic wave functions in the graphene surface.

In order to analyze the interference patterns, we give 
comparison with the calculation of the model for the graphene 
plane.  Here, we use the continuum $\kp$ model \cite{10}.  
The electron density is calculated with including a static 
potential which has a functional form of the linear decrease 
from top to the bottom along the surface of the graphene necktie. 
The peak positions of the electron density in the long direction 
of the graphene necktie of Fig. 5 are plotted in Fig. 6, and 
comparison with the theoretical results is given.  The decrease 
of the oscillation period fairly agrees with the experiments. 
However, the slight decrease of the experimental corrugation 
cannot be reproduced by the theoretical result because of 
neglecting an effect of a tip-apex shape of STM on the observed 
corrugatrion amplitude. The fitting gives the parameter of the
potential gradient $6.49 \times 10^{-3}$ eV/nm. The total potential 
variation over the distance 200 nm becomes 1.3 eV.  Such 
magnitude of the potential change would survive under thermal 
lattice fluctuations and can really exist in experiments.  
The present result by no means implies that the wave functions 
observed with superperiodic amplitudes are of the electrons 
which have energy levels of the graphene plane.

\section{Summary}

First, antiferromagnetism in the stacked nanographite has been 
investigated with the Hubbard-type model.  The A-B 
stacking is favorable for the hexagonal nanographite 
with zigzag edges, in order that magnetism appears.  
We have also found that the open shell electronic 
structures can be origins of the decreasing magnetic 
moment with adsorption of molecules.

Next, possible charge-separated states have been 
considered using the extended Hubbard model with nearest-neighbor 
interactions.  The charge-polarized state could appear, 
when a static electric field is applied in the graphene 
plane for example.

Finally, we have characterized theoretically the superperiodic 
patterns in a nanographene sheet observed by STM.
We have adopted the $\kp$ model for the description of 
the electronic structures of the graphite.  The calculated 
electron density has the property that the oscillation 
period decreases spatially while the amplitude remains constant.  
The magnitude of the static potential seems reasonable.  
It turned out that the long distance oscillations come from 
electrons with the band structures of the two dimensional 
graphene sheet.

\pagebreak

\noindent
{\bf Figure Captions}

\mbox{}

\noindent
Fig. 1. (a) A-B stacked hexagonal nanographite with 
zigzag edges. (b) The absolute magnitude of the total magnetic
moment per layer as a function of $t_1$.  The onsite interaction
is varied within $1.8t$ (closed squares) $\leq U \leq 2.3t$ 
(open triangles).  The interval of $U$ between the series of 
the plots is $\Delta U = 0.1t$.

\mbox{}

\noindent
Fig. 2. The absolute magnitude of the total magnetic
moment per layer as a function of $t_1$ for the system with
a site potential $E_s = -2 t$, (a) at the site A and (b) at 
the site B.  The site positions are displayed in Fig. 1 (a).  
In (a), the onsite interaction is varied within $0.6t$
(closed squares) $\leq U \leq 1.8t$ (closed triangles)
with the interval $\Delta U = 0.3t$.  In (b), it is varied 
within $1.0t$ (closed squares) $\leq U \leq 2.0t$ (closed triangles)
with the interval $\Delta U = 0.25t$.

\mbox{}

\noindent
Fig. 3. (a) Charge density distribution of the 
charge-separated (CS) state, and (b) the spin density 
distribution of the localized magnetic (LM) state on a 
zigzag ribbon with $4{\times}20$ carbon atoms, where
$\circ$ means positive charge (spin) densities, and $\bullet$ 
indicates negative charge (spin) densities.  The radius of 
each circle denotes the magnitude of the density.

\mbox{}

\noindent
Fig. 4. The phase diagram in the parameter space of $U$ 
and $V$ for a zigzag ribbon with $4{\times}40$ carbon atoms. 
The solid curve is the boundary between the CS and LM states,
interpolating the numerical data ($\bullet$). The inset shows 
the phase diagram in reduced scales.  The dashed line denotes 
the phase boundary in the strong correlation limit.

\mbox{}

\noindent
Fig. 5. STM image of the superperiodic pattern observed
on a necktie shaped graphene plate on HOPG substrate.

\mbox{}

\noindent
Fig. 6. Comparison for the electron wave patterns by 
STM and the $\kp$ model.  Experimental peak positions 
along the perpendicular direction of Fig. 5 are plotted 
by diamonds.  The results of the fitting by the long 
distance envelope functional form derived from the 
$\kp$ model are shown by squares.

\pagebreak

\begin{figure}
\begin{center}
\includegraphics{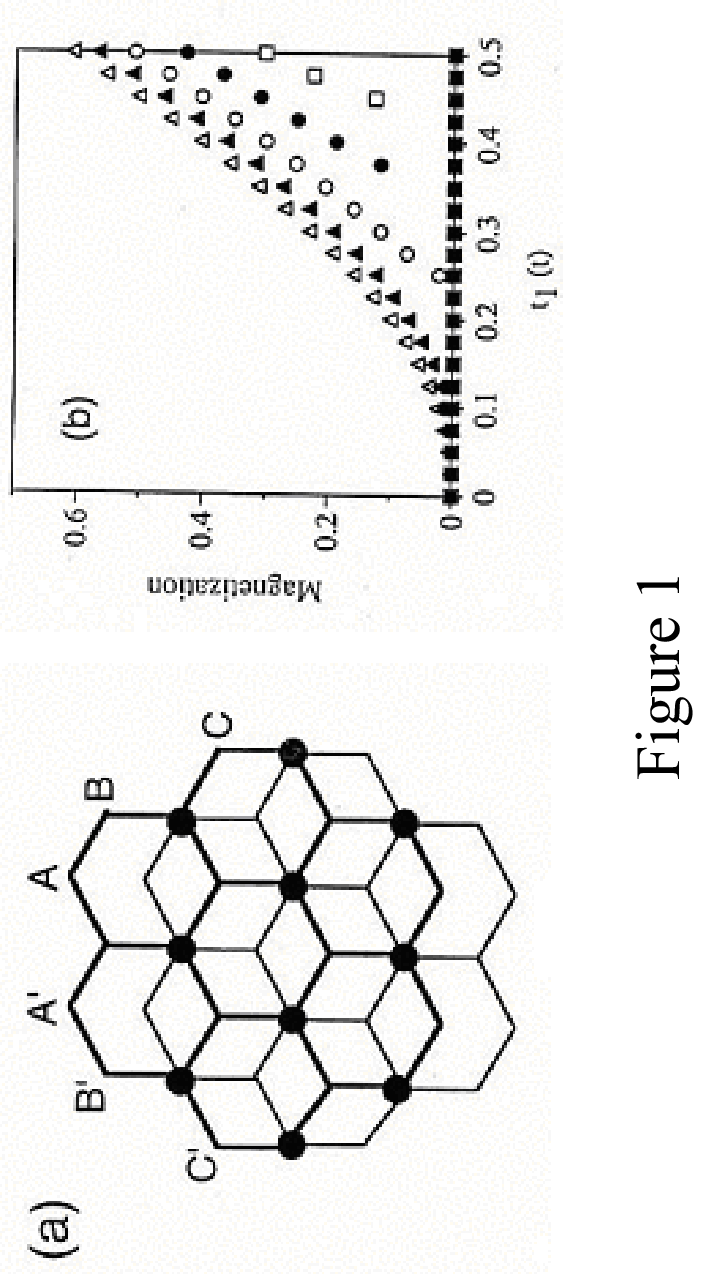}
\end{center}
\caption{~}
\label{Fig1}
\end{figure}

\pagebreak

\begin{figure}
\begin{center}
\includegraphics{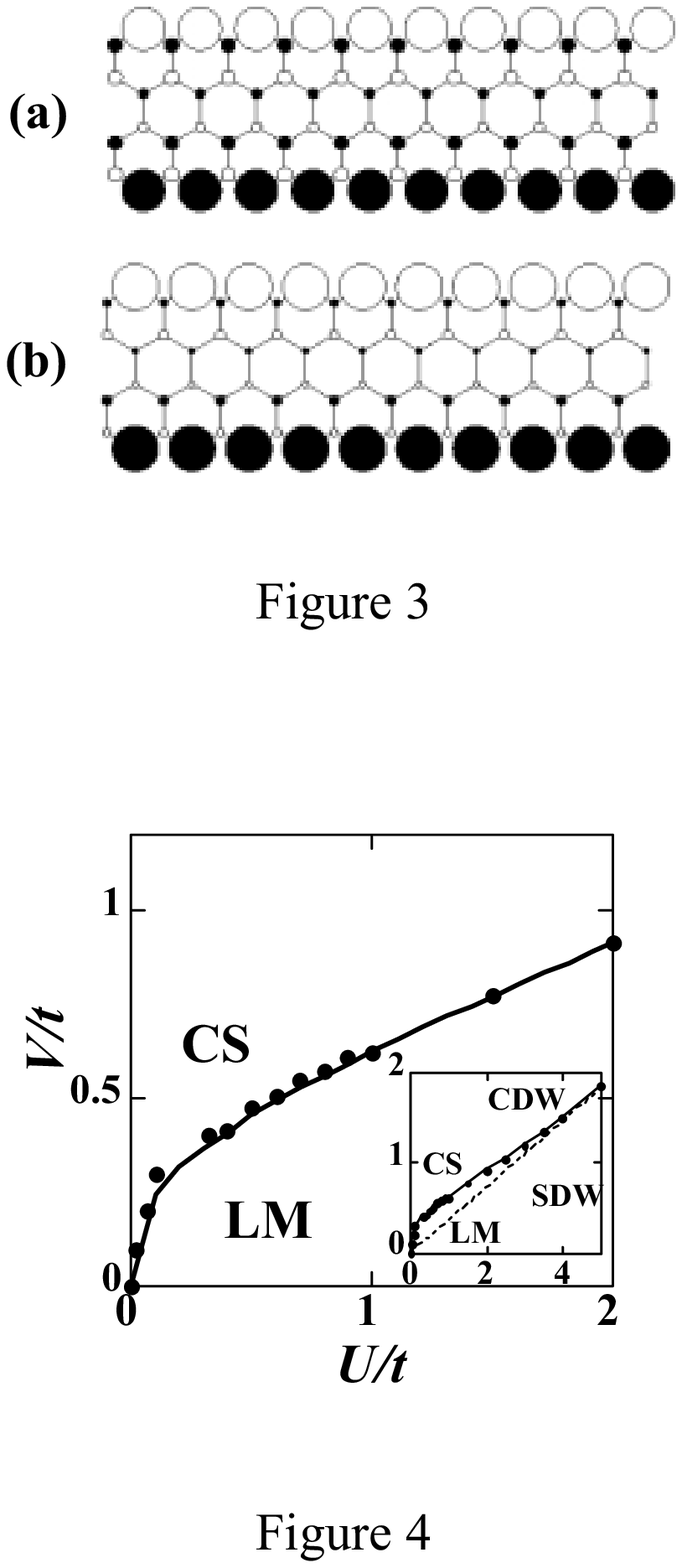}
\end{center}
\caption{~}
\label{Fig3}
\end{figure}

\pagebreak

\begin{figure}
\begin{center}
\includegraphics{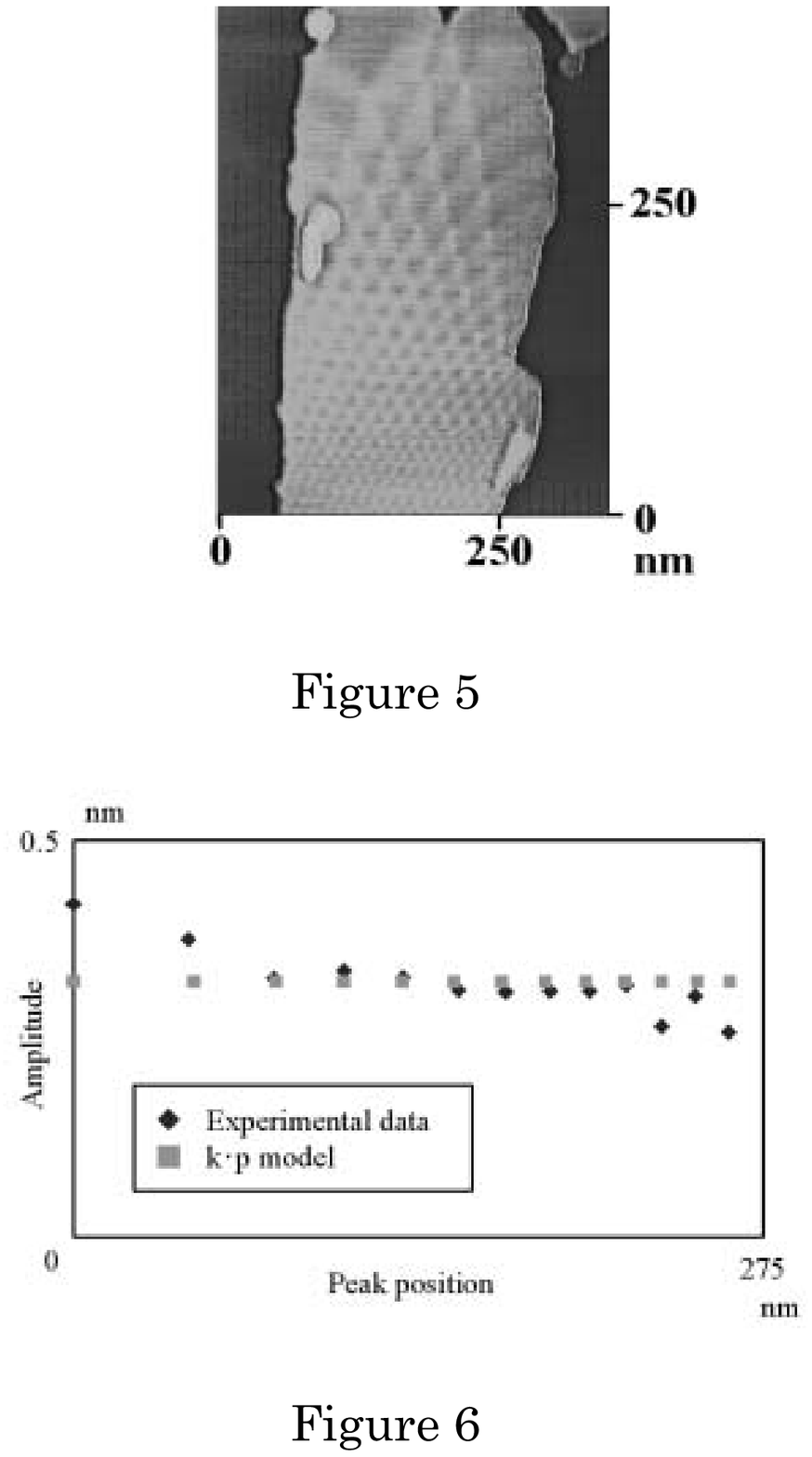}
\end{center}
\caption{~}
\label{Fig5}
\end{figure}

\end{document}